\documentclass[structabstract]{aa}
\usepackage{graphicx} 
\usepackage{txfonts} 

\begin{document}

\title{Synthesis of C--rich dust in CO nova ourbursts}

\author{Jordi Jos\'e \inst{1,2}
   \and Ghina M. Halabi \inst{3}
   \and Mounib F. El Eid \inst{4} }

\offprints{J. Jos\'e}

 \institute{Departament de F\'\i sica, EUETIB,
            Universitat Polit\`ecnica de Catalunya, 
            c/Comte d'Urgell 187, 
            E-08036 Barcelona, 
            Spain\
            \and 
            Institut d'Estudis Espacials de Catalunya, 
            c/Gran Capit\`a 2-4, 
            Ed. Nexus-201, 
            E-08034 Barcelona, 
            Spain\
            \and
            Institute of Astronomy,
            University of Cambridge,
            Madingley Road,
            Cambridge CB3 0HA,
            United Kingdom
            \and
            Department of Physics, 
            American University of Beirut, 
            Bliss St. 11-0236, 
            Beirut 1107 2020, Lebanon\\
      \email{jordi.jose@upc.edu}}
       
\date{\today}

\abstract{Classical novae are thermonuclear explosions that take place in the envelopes of accreting white dwarfs in stellar binary systems. The material transferred onto the white dwarf piles up under degenerate conditions, driving a thermonuclear runaway. In those outbursts, about $10^{-7} - 10^{-3}$ $M_\odot$, enriched in CNO and, sometimes, other intermediate-mass elements (e.g., Ne, Na, Mg, or Al, for ONe novae) are ejected into the interstellar medium. 
The large concentrations of metals spectroscopically inferred in the nova ejecta reveal that the (solar-like) material transferred from the secondary mixes with the outermost layers of the underlying white dwarf.}
         {Most theoretical models of nova outbursts reported to date yield, on average, outflows characterized by O $>$ C, 
from which only oxidized condensates (e.g, O-rich grains) would be expected, 
in principle.}
         {To specifically address  whether CO novae can actually produce C-rich dust, six different hydrodynamic nova models have been evolved, from accretion to the expansion and ejection stages, with different choices for the composition of the substrate with which the solar-like accreted material  mixes.  
Updated chemical profiles inside the H-exhausted core have been used, based on stellar evolution calculations for a progenitor of 8 $M_\odot$ through H and He-burning phases. }
{We show that these profiles lead to C-rich ejecta after the nova outburst. This extends the possible contribution of
novae to the  inventory of presolar grains identified in meteorites, particularly in a number of carbonaceous phases (i.e., nanodiamonds,
silicon carbides and graphites).}
         {}


\titlerunning{Synthesis of C--rich dust in CO novae} 
\authorrunning{J. Jos\'e et al.} 

\maketitle

\section{Introduction}
Classical novae are prolific dust producers. Indeed, 
infrared and ultraviolet observations 
have unambiguously revealed a number of dust forming episodes in
the ejected shells accompanying nova outbursts 
(Evans 1990, Evans \& Rawlings 2008, Gehrz et al. 1998, 
Gehrz 2002, Gehrz 2008, Shore et al. 1994). 
As first suggested by Stratton \& Manning (1939), on the 
 occasion of nova DQ Her,
dust formation is revealed by a rise in infrared emission that
occurs simultaneously to a decline in the optical lightcurve, 
several months after peak luminosity. 

Such dust forming episodes are driven by the
fast decline in temperature and density in the nova outflows,
which allows the plasma to recombine and subsequently form 
 molecules.  Grain (dust) formation is a complex process dictated by the local 
thermodynamical and physicochemical conditions of the environment,
  requiring moderately low temperatures, below 1500--2000 K, 
and particle densities $\geq 10^8$ cm$^{-3}$. 
Roughly speaking, it can be qualitatively described as a two-stage process:
grain nucleation and growth to macroscopic size 
(Sedlmayr 1994, Lodders 2003, Andersen 2011).
In the former, a handful of molecules assemble into small groups or
{\it clusters}, which by effect of a suite of chemical processes grow to a 
critical size. The carbon monoxide (CO) molecule plays an important role in this
process.  While most heteronuclear, diatomic molecules have dissociation energies 
around 3--5 eV, the CO molecule  is characterized by a striking bond energy of 11.2 eV. Therefore, 
it is a very stable molecule, and can only be dissociated by high-energy photons.  
In the absence of intense radiation fields,
condensation in an O-rich plasma (i.e.,  O $>$ C) results in nearly all carbon
atoms locked in the form of the very stable CO molecules. 
As a result, formation of C-rich grains, such as silicon carbides (SiC), graphites, or nanodiamonds,
is unlikely. 
Conversely, in a C-rich environment (C $>$ O), all oxygen gets trapped in CO molecules
and oxidized compounds (e.g., corundum, spinel, hibonite, enstatite...) cannot form, in principle.

 Infrared measurements of a number of novae have actually revealed the presence of
 C-rich dust (Aql 1995, V838 Her 1991, PW Vul 1984),
 SiC (Aql 1982, V842 Cen 1986), hydrocarbons (V842 Cen 1986, V705 Cas 1993), 
or SiO$_2$ (V1370 Aql 1982, V705 Cas 1993) in the ejecta.
Moreover, sequential formation of all those types of dust 
has also been reported for some remarkable novae,  such as 
 V842 Cen 1986 (Gehrz et al. 1990) and 
QV Vul 1987 (Gehrz et al.  1998). 
In fact, it has been suggested that 
novae may have contributed to the different presolar grain populations
isolated from meteorites\footnote{See also Pepin et al. 2011, for
the possible nova origin of anomalous interplanetary dust particles
collected from comets Grigg-Skjellerup and Tempel-Tuttle.} (Clayton \& Hoyle 1976, Amari et al. 2001, Amari 2002, 
Jos\'e et al. 2004, Haenecour et al. 2016). 
Surprisingly, most theoretical models of nova outbursts (even those that
rely on a CO-rich white dwarf as the underlying compact object that hosts the
outburst) yield, on average, outflows characterized by O $>$ C, 
from which only oxidized condensates would be expected\footnote{See
Gehrz et al. 1998, Gehrz 2008, 
and Jos\'e \& Shore 2008, for examples of O $>$ C ejecta observed 
in CO novae, at different stages of the outburst.}.
Equilibrium condensation sequences have suggested, however,
that the presence of large amounts of intermediate-mass elements, such as Al, Ca, Mg, or Si,
 may dramatically alter the condensation process, allowing the formation of C-rich dust
even in a slightly O-rich environment (Jos\'e et al. 2004).  
Moreover, it is likely expected that dust condensation in a nova environment proceeds kinetically 
rather than at equilibrium, because of the strong radiation emitted by the 
underlying white dwarf (see, e.g., Shore \& Gehrz 2004). 
Both effects reduce the significance of the CO molecule in the process
of dust formation. 
But whether the nova ejecta are actually C-- or O--rich  has not been systematically addressed in
the framework of hydrodynamic simulations. 
Indeed, nova models have traditionally assumed that
the composition of the underlying white dwarf star that hosts the outbursts, has similar mass
fractions of $^{12}$C and $^{16}$O, when low-mass, CO-rich white dwarfs are considered
(see, however, Kovetz \& Prialnik 1997, for a study of the effect of the white dwarf composition
on the basis of pure-C, pure-O and standard, C/O = 1 white dwarf models). 

In this paper, a series of nova models are presented, with state-of-the-art chemical
profiles based on new evolutionary sequences of intermediate-mass stars that lead to 
the formation of white dwarfs with outer layers characterized by C $>$ O.   
An outline of the method of computation and input physics used in this paper is given in
Section 2. The impact 
of  the  abundance profiles of  the underlying  white dwarf  on  the nucleosynthesis
accompanying classical nova outbursts is presented in Section 3. Finally, 
the most relevant conclusions of this paper are summarized in Section 4.

\section{Input physics and initial setup}

The nova models reported in this paper have been  computed with the one-dimensional,
implicit, Lagrangian, hydrodynamic code {\tt SHIVA}, extensively used in the
the modeling of stellar explosions (e.g., classical novae, X-ray bursts). 
The code solves the standard set of differential equations of 
stellar evolution: conservation of mass, momentum, and
energy, and energy transport by radiation and (time-dependent) convection.
The equation of state
includes contributions from the (degenerate) electron gas, the ion plasma, and radiation. 
Coulomb corrections to the electron pressure are taken into account. 
Radiative and conductive opacities are considered. The code is linked to a nuclear reaction 
network that contains $\sim 120$ nuclear species, from H to $^{48}$Ti,
connected through 630 nuclear interactions, with updated {\tt STARLIB} 
rates (Sallaska et al. 2013).
Additional details on the {\tt SHIVA} code can be 
found in Jos\'e \& Hernanz (1998) and Jos\'e (2016).

\begin{figure}[h!]
\includegraphics[scale=0.30]{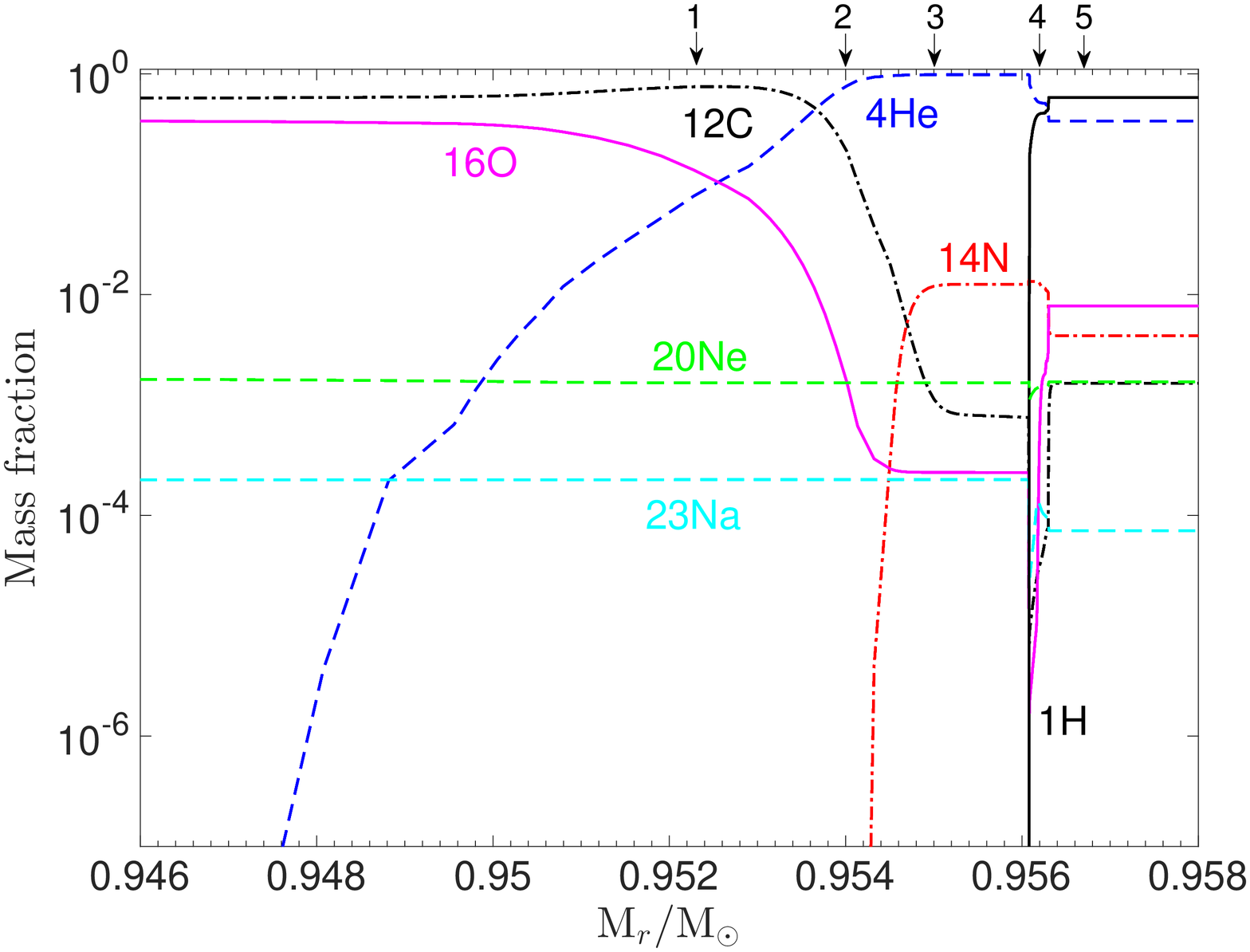}
\caption{Chemical profiles after a series of thermal pulses, following 
completion of central He-burning, for an 8 $M_\odot$ star, 
computed with the {\tt HYADES} code.}
\end{figure}

The nova models computed in this work assume a $\sim$ 1.0 $M_\odot$ CO white dwarf, resulting from the evolution of an 8 $M_\odot$ progenitor star, self-consistently computed throughout the successive H- and He-burning stages with the {\tt HYADES} code (see 
Halabi, El Eid \& Champagne 2012, and references therein). 
 The data base of the Nuclear Astrophysics Compilation of Reaction Rates 
(hereafter, NACRE; Angulo et al. 1999) was used, in particular for 
the triple-$\alpha$ and $^{12}$C($\alpha$, 
$\gamma$) reactions.
The chemical structure of the resulting CO-rich white dwarf is shown in Fig. 1.  Qualitatively similar chemical profiles have been obtained for 
 the remnants of 6 $M_\odot$ and 7 $M_\odot$ progenitor stars.

Most of the simulations of CO (non-neon) novae performed to date have traditionally 
assumed that the outer layers of the white dwarf hosting the outburst have equal
mass fractions of $^{12}$C and $^{16}$O (some include a marginal amount of 
leftover $^{22}$Ne, at a level of 1\% by mass, but maintaining X($^{12}$C)/X($^{16}$O) = 1). However,
a closeup look at Fig. 1 reveals instead a chemical abundance pattern dominated
by the presence of $^{12}$C over $^{16}$O in the outer white dwarf layers (with the exception of the outermost, unburnt H-rich layers). 

In the simulations discussed in Sect. 3, the white dwarf is assumed to accrete solar composition material ($Z \sim 0.02$) from a low-mass stellar companion,
at a characteristic mass-accretion rate of $\dot M \sim 2 \times 10^{-10}$ $M_\odot$ yr$^{-1}$. 
The accreted material is mixed with material dredged-up from the underlying white dwarf  (a mixture with 75\% solar composition plus 25\% outer white dwarf 
material has been adopted in all the calculations reported in this paper; 
 see Kelly et al. 2013, for a recent reanalysis of mixing fractions in 
novae, supporting a 25\% white dwarf contribution. 
See also Casanova et al. 2010, 2011a,b, for 2-D and 3-D hydrodynamic simulations of mixing at the core-envelope interface during nova outbursts). 

To specifically investigate the role played by the composition of the underlying white dwarf, six different nova models have been evolved. Models 1 to 5 are identical, except for the composition of the substrate with which the solar-like accreted material mixes up. To this end, 5 different locations in the outer white dwarf layers have been considered (see points indicated by arrows in Fig. 1). For comparison, an additional model (Model 6) has been evolved under the traditional assumption of equal mass fractions of $^{12}$C and $^{16}$O in the outer white dwarfs layers.

\begin{table}
\caption{Initial composition of the accreted envelopes.}
\label{table:1}      
\centering                                  
\begin{tabular}{c c c c c}          
\hline
\hline        
Model & $^1$H & $^{12}$C& $^{16}$O& CNO\\
\hline                                 
  1   & 0.529  & 0.193  & 4.11-2  & 0.235\\
  2   & 0.529  & 5.21-2 & 7.95-3  & 6.10-2\\
  3   & 0.529  & 2.49-3 & 7.61-3  & 1.42-2\\
  4   & 0.638  & 2.28-3 & 7.76-3  & 1.40-2\\
  5   & 0.682  & 2.67-3 & 9.51-3  & 1.42-2\\
  6   & 0.529  & 0.126  & 0.131   & 0.258\\
\hline
\hline      
\end{tabular}
\end{table}

\begin{table*}
\caption{Main characteristics of the nova outbursts.}
\label{table:2}      
\centering                                  
\begin{tabular}{c c c c c c c c}          
\hline
\hline        
Model & [X($^{12}$C) $\times$ X($^1$H)]$_{\rm ini}$& 
P$_{\rm base,max}$& M$_{\rm acc}$& T$_{\rm base,max}$& M$_{\rm ejec}$&  V$_{\rm mean}$& K$_{\rm mean}$\\
   &   & ($10^{19}$ dyn cm$^{-2}$)& ($10^{-5}$ $M_\odot$)& ($10^8$ K)& ($10^{-5}$ $M_\odot$)&   (km s$^{-1}$)& ($10^{44}$ ergs)\\
\hline                                 
  1   & 0.102 & 0.857 & 3.94 & 1.72 & 3.19 & 1540 & 9.25\\
  2   & 2.76-2& 1.09  & 5.05 & 1.78 & 4.08 & 1210 & 7.26\\
  3   & 1.32-3& 1.52  & 7.07 & 1.89 & 5.71 & 1060 & 8.18\\
  4   & 1.46-3& 1.43  & 6.72 & 1.83 & 5.43 & 983  & 6.81\\
  5   & 1.82-3& 1.37  & 6.43 & 1.80 & 5.20 & 954  & 6.19\\
  6   & 6.67-2& 0.901 & 4.15 & 1.73 & 3.35 & 1400 & 7.90\\
\hline
\hline      
\end{tabular}
\end{table*}

\begin{table*}
\caption{Mean composition of the ejecta (CNO-group nuclei).}
\label{table:3}      
\centering                                  
\begin{tabular}{c c c c c c c c c}          
\hline
\hline        
Model & $^{12}$C& $^{13}$C& $^{14}$N& $^{15}$N& $^{16}$O& $^{17}$O& $^{18}$O& Ejecta\\
\hline                                 
  1   & 2.90-2  & 7.56-2  & 1.06-1  & 8.42-3  & 3.85-2  & 8.36-4  & 5.27-7  & C-rich\\
  2   & 8.89-3  & 1.43-2  & 3.56-2  & 2.73-3  & 5.62-3  & 1.46-4  & 9.23-8  & C-rich\\
  3   & 2.04-3  & 3.02-3  & 7.21-3  & 5.26-4  & 3.07-4  & 9.46-6  & 6.52-9  & C-rich\\
  4   & 1.79-3  & 2.66-3  & 7.73-3  & 4.04-4  & 3.92-4  & 1.01-5  & 7.19-9  & C-rich\\
  5   & 1.63-3  & 2.47-3  & 8.08-3  & 3.44-4  & 5.09-4  & 1.14-5  & 8.24-9  & C-rich\\
  6   & 2.05-2  & 4.17-2  & 8.20-2  & 7.26-3  & 1.17-1  & 2.75-3  & 1.72-6  & O-rich\\
\hline
\hline      
\end{tabular}
\end{table*}

\section{Results}

\subsection{Model 1}
The importance of the initial C/O ratio in the outer
layers of the underlying white dwarf, and especifically, the initial 
$^{12}$C mass fraction in the envelope, will be described in the framework of
 the evolution of Model 1. In this particular model, the solar material transferred from the secondary is assumed to mix with a C-rich substrate, whose composition is indicated in the upper part of Fig. 1 as point 1 (see also Table 1). 
The early accretion phase is dominated both by the pp chains
(mainly through $^1$H(p, e$^+$ $\nu$)$^2$H) as well as by the CNO-cycle reaction $^{12}$C(p, $\gamma$)$^{13}$N, followed by 
$^{13}$N($\beta^+$)$^{13}$C. 
When the temperature at the envelope base reaches
T$_{\rm base} = 2.4 \times 10^7$ K, the nuclear timescale becomes shorter than the
accretion timescale, and accretion becomes no longer relevant.  
The rate of nuclear energy generation reaches $\sim 10^8$ 
ergs g$^{-1}$ s$^{-1}$ at the innermost envelope shell.
The total mass piled up on top of the white dwarf at the end of this accretion atage (which lasts $\sim 1.7 \times 10^5$ yr) is $\sim 3.9 \times 10^{-5}$ $M_\odot$ (see Table 2, for details). 

Shortly afterwards, degeneracy is lifted at the envelope base, 
so that pressure  becomes 
sensitive to temperature. Consequently, the envelope begins to expand, after achieving a maximum pressure
of P$_{\rm max,base} = 8.6 \times 10^{18}$ dyn cm$^{-2}$ 
and a maximum density of
$\rho_{\rm max,base} = 3.6 \times 10^3$ g cm$^{-3}$. 
The beginning of the dynamic stage (i.e., a thermonuclear runaway) is accompanied by the 
development of superadiabatic gradients at the innermost layers
of the envelope. This spawns the onset of convection, 
which progressively expands throughout the envelope. 
At T$_{\rm base} = 5 \times 10^7$ K, the nuclear
activity is fully dominated by the cold CNO cycle,
mainly through $^{12}$C(p, $\gamma$)$^{13}$N($\beta^+$)$^{13}$C(p, 
$\gamma$)$^{14}$N, while
 no significant activity in the NeNa and MgAl-mass region is noticed. A similar behavior is found at $10^8$ K, when convection 
has already extended throughout the entire envelope, 
with a characteristic turnover time of about 1 s.
However, the dominant 
$^{12}$C(p, $\gamma$)$^{13}$N($\beta^+$)$^{13}$C(p, 
$\gamma$)$^{14}$N reactions are now significantly
supplemented by 
$^{13}$N(p, $\gamma$)$^{14}$O($\beta^+$)$^{14}$N(p, $\gamma$)$^{15}$O 
and $^{16}$O(p, $\gamma$)$^{17}$F,
with still a marginal contribution from reactions of the NeNa and MgAl mass-regions (mainly, $^{24}$Mg(p, $\gamma$)$^{25}$Al).
The star has spent about $1.2 \times 10^6$ s raising its temperature 
at the innermost envelope layers from $3 \times 10^7$ K
to $10^8$ K.

About 100 s later, 
when the temperature reaches T$_{\rm base} = 1.66 \times 10^8$ K, 
the star 
achieves a maximum rate of nuclear energy generation of 
$3 \times 10^{16}$ ergs g$^{-1}$ s$^{-1}$. 
Forty seconds later, the envelope base achieves a peak
temperature of T$_{\rm base,max} = 1.72 \times 10^8$ K.
At this stage, nuclear energy 
is mostly released through the hot CNO cycle reactions (HCNO1)
$^{13}$N(p, $\gamma$)$^{14}$O($\beta^+$)$^{14}$N, 
supplemented by 
$^{12}$C(p, $\gamma$)$^{13}$N($\beta^+$)$^{13}$C(p, 
$\gamma$)$^{14}$N(p, $\gamma$)$^{15}$O($\beta^+$)$^{15}$N(p,
 $\alpha$)$^{12}$C, and
$^{16}$O(p, $\gamma$)$^{17}$F($\beta^+$)$^{17}$O(p, $\alpha$)$^{14}$N.
Secondary nuclear activity involves 
a series of proton-capture reactions
and $\beta^+$ decays of a handful of species in the
NeNa-MgAl mass regions (e.g., $^{23}$Na, $^{24,25,26}$Mg, $^{25}$Al,
and both the ground and first isomeric states of $^{26}$Al).
A small leakage from the MgAl-mass
region is driven by proton captures on $^{27}$Si and $^{27}$Al.

A fraction of the nuclear energy released during the event
is transformed into kinetic energy, which powers the ejection 
of $3.2 \times 10^{-5}$ $M_\odot$
 of nuclear-processed material, with a mean velocity of
1540 km s$^{-1}$. As confirmed by 
the large concentrations of hydrogen and carbon in the ejecta
(see Table 3), 
the runaway is halted
by envelope expansion, rather than by fuel consumption.

\subsection{Models 2 to 6}

As shown by Shara (1981) and Fujimoto (1982), the key parameter that 
determines the strength of a nova outburst is the pressure achieved at
the core-envelope interface, P$_{\rm base}$, a measure of the overall 
pressure exerted by the layers overlying the ignition shell: 
\begin{equation}
P_{\rm base} = \frac{G \, M_{\rm wd}}{4 \pi R_{\rm wd}^4} M_{\rm acc}
\end{equation}
In Equation 1, M$_{\rm wd}$ and R$_{\rm wd}$ are the mass and radius 
of the white dwarf hosting the eplosion, and M$_{\rm acc}$ is the mass of the 
accreted envelope. To drive mass ejection,  pressures in the range 
P$_{\rm base} \sim 10^{19} - 10^{20}$ dyn cm$^{-2}$ 
are required at the base of the 
envelope, the exact value depending on the chemical composition 
(Fujimoto 1982, MacDonald 1983). For a given pressure, Equation 1 
reveals that the mass of the accreted envelope depends only on the 
mass of the underlying white dwarf star, because of the relationship 
between stellar mass and radius. As shown in Table 2, the larger the 
mass of the accreted envelope, the larger the pressure at its base.
 This translates into a more violent outburst, characterized by a 
larger peak temperature, T$_{\rm base,max}$ (Table 2). 

The analysis reported for Model 1 revealed that the single, most important 
reaction during the early stages of a nova outburst\footnote{Specifically, 
for CO novae. See Shen and Bildsten (2009),  for ignition conditions 
in C-poor envelopes.} is $^{12}$C(p, $\gamma$). 
The rate at which these two species interact per second and per unit volume 
can, in general, be expressed as (see, e.g., Iliadis 2015, for details):
\begin{equation}
r_{\rm HC} = n_{\rm H} n_{\rm C} \langle \sigma v \rangle_{\rm HC}
\end{equation}
where $n_{\rm H}$ and $n_{\rm C}$ denote the number densities of $^1$H and $^{12}$C in
the stellar plasma, respectively, and $\langle \sigma v \rangle_{\rm HC}$ is
the reaction rate per particle pair. Equation 2 reveals that the rate
of interaction is proportional to the mass fractions of the interacting
species. Therefore, the most relevant quantity to determine the strength
of the outburst is not the amount of $^{12}$C, as it is frequently stated, 
but the product [X($^1$H) $\times$ X($^{12}$C)]$_{\rm 
ini}$ at the beginning of the accretion stage.
A
reduction in this quantity delays ignition, since less nuclear reactions
occur, and, hence, less energy is released. This translates into a rise 
in the characteristic
timescale for accretion, which subsequently leads to a larger accreted mass, 
larger pressure at the base, and in the end, a more violent outburst\footnote{See Jos\'e et al. (2007), for
a study of  nova outbursts in metal-poor, primordial-like binaries.}.
Indeed, in the models reported in this work,
a reduction in
[X($^1$H) $\times$ X($^{12}$C)]$_{\rm ini}$ by a factor of $\sim 80$
(see, e.g., Models 1 and 3 in Table 2)
translates into an increase by a factor $\sim 2$ in the maximum pressure
attained at the base of the envelope, as well as in the amount of mass
accreted and ejected (with a smaller effect as well on the peak temperature
achieved in the outburst). 

\subsection{Nucleosynthesis and implications for dust formation}

The final, mass-averaged composition of the ejecta accompanying the
set of nova outbursts calculated in this work is summarized, for CNO-group
species, in Table 3.

In Model 1, the ejecta is dominated by high concentrations of H
(0.498, by mass), $^4$He (0.240), and $^{14}$N (0.106). This is followed
by $^{13}$C, $^{16}$O, $^{12}$C, and other CNO-group nuclei.
A similar pattern is obtained in Models 2 to 5,  except that in the
latter, 
 both
$^{12,13}$C are individually more abundant than any of the stable O isotopes,
$^{16,17,18}$O. It is worth noting that Models 1 to 5 yield C-rich ejecta.
Model 6, however,
 presents an important deviation: while H and $^4$He still present
the largest mass fractions in the ejecta, the third most abundant species
 is now, by far, $^{16}$O. In fact, the large concentration  of this isotope
 (X($^{16}$O) = 0.117, by mass) results in O-rich ejecta.

While
the use of state-of-the-art chemical profiles for 
CO-rich white dwarfs hosting nova outbursts 
has a limited impact on the
dynamics of the nova outburst (i.e., mass ejected, peak temperature, 
nucleoynthesis) it has a dramatic effect on the properties of the
dust expected
to condense in those environments. 
To date, CO novae have been  traditionally ruled out as 
likely progenitors of C-rich grains, on the basis of the
O-rich ejecta 
 obtained when
equal mass fractions of $^{12}$C and $^{16}$O are adopted (see, e.g.,
Jos\'e et al. 2004).
In sharp contrast, the 
 models reported in this paper show, however, that
the implementation of realistic chemical profiles leads 
to C-rich ejecta, instead.
This opens up new possibilities for the contribution of novae to 
the 
inventory of presolar grains identified in meteorites,
particularly in a number of carbonaceous phases (i.e., nanodiamonds,
silicon carbides and graphites).

\section{Discussion}
As shown in Section 3, the use of state-of-the-art chemical  profiles for the white dwarf that hosts a CO nova outburst turns out to be critical for the production of C-rich ejecta.
Even though realistic white dwarf models, with similar profiles, have been previously reported in the literature (see, e.g., Salaris et al. 1997, Karakas and Lugaro 2016), it is worth analyzing the possible role played by nuclear uncertainties in this regard, both during the previous stages of the evolution of the progenitor and during the nova outburst itself.

\subsection{Uncertainties in nova nucleosynthesis}
A significant fraction of the nuclear reactions 
of interest for nova nucleosynthesis that involve 
stable nuclei (mostly proton-induced reactions)
have already been measured directly in the laboratory
(Jos\'e, Hernanz \& Iliadis 2006).

For the case of CO novae, in which
the main nuclear path does not extend significantly beyond the
CNOF mass region, current reaction rate uncertainties 
(i.e., $\leq$ 30\%) do not dramatically affect model predictions. 
Indeed, after a number of 
improvements on the $^{17}$O(p, $\alpha$) and 
$^{17}$O(p, $\gamma$) rates,  
the main source of uncertainty in this region is mostly dominated by
the challenging reaction $^{18}$F(p, $\alpha$) which,
under explosive conditions, drives the main nuclear path
down to $^{15}$O in both hot CNO cycles 2 and 3 (see, e.g., 
Iliadis 2015, Jos\'e 2016).
It is, however, worth noting that
 the uncertainty associated  with $^{18}$F(p, $\alpha$) has an 
impact on the expected O yields for ONe novae, but not
for CO novae (Iliadis et al. 2002).

All in all, we conclude that the CO nova nucleosynthesis predictions  
reported in this work are not affected by current reaction-rate 
uncertainties.

\subsection{Uncertainties in the evolution of the progenitor star}
The evolution of a star, throughout the H- and He-burning stages, up to the formation of a CO-rich core, may be affected by uncertainties in 
two key nuclear reaction rates.

The first is the triple-alpha reaction (hereafter, $3\alpha$). 
Models presented in this work rely on chemical profiles for the white dwarf 
based on stellar evolution computed with the {\tt  HYADES} code, 
for which the NACRE $3\alpha$ rate was adopted (Angulo et al. 1999). 
Since the NACRE compilation, several prescriptions for this rate have been published.
Through a measurement of the inverse process, $^{12}$C $\rightarrow$ 3 $^4$He, 
Fynbo et al. (2005) determined a new $3\alpha$ rate for temperatures  
ranging between $10^7$ K and $10^{10}$ K. The new rate was found to be higher than 
the NACRE rate for 
 temperatures below $\sim 5 \times 10^7$K, while much smaller 
than NACRE above $10^9$ K. The potential effect of the Fynbo et al. 
3$\alpha$ rate has been checked by means of new evolutionary sequences of 
an 8 $M_\odot$ star computed with the {\tt HYADES} code throughout the 
He-burning stage. Oxygen becomes higher than carbon in the inner regions 
of the star, but as in the chemical profiles adopted in this work, the outer 
layers still remain C-rich. 
More recently, Ogata et al. (2009) have reported on a new rate based 
 on quantum mechanical calculations of the
three-body Schr\"odinger equation. Around $10^7$ K, the rate exhibited an increase by 20 orders of magnitude 
 over the standard NACRE rate. It is however worth noting that this rate has been refuted by more recent calculations, and differences with respect to NACRE at low temperatures are now reduced to less than an order of magnitude (see, e.g., Ishikawa 2013, Akahori, Funaki, \& Yabana 2015, for details).

The second reaction is $^{12}$C($\alpha$, $\gamma$)$^{16}$O. 
 It is a key reaction determining the final C/O ratio wherever He-burning 
takes place,  and dominates over the 
triple-$\alpha$ reaction when the helium mass fraction drops below about 0.1. 
The rate of this reaction is dominated by several subthreshold resonances and  still remains uncertain, even though it has been re-evaluated several times
 since NACRE (e.g., Katsuma et al. 2012, Xu et al. 2013). The Katsuma et al. rate is higher than the NACRE rate for temperatures above $\geq 3 \times 10^8$ K, while the Xu et al. rate exceeds the NACRE rate above $\geq 5 \times 10^8$ K. 
To estimate the potential effect of the He-burning rates on the results 
reported in this paper, new evolutionary sequences have been computed 
with the {\tt HYADES} code using the Fynbo et al. and the Xu et
al. prescriptions for the 3$\alpha$ and 
 $^{12}$C($\alpha$, $\gamma$)$^{16}$O
rates, respectively. The implementation of these rates results
in an important increase of the oxygen abundance in the inner layers 
of the star (up to  $\sim 0.75$ $M_\odot$,
at the end of He-burning), however, the outer layers are still C-rich.

To investigate how our conclusions are affected by the latest He-burning 
 rate determinations, we recomputed an 8 $M_\odot$ model using the Fynbo et al. (2005) rate for the $3\alpha$ reaction and Xu et al. (2013) for the $^{12}$C($\alpha$, 
 $\gamma$) reaction. We found out that the outer core remains C-rich. A 6 $M_\odot$ model by Karakas 
\& Lugaro (2016) using the Fynbo et al. (2005) rate and the Xu et al. (2013) rate for the $3\alpha$ and the $^{12}$C + $\alpha$ reactions, respectively, 
yields similar conclusions.

A very recent work by Fields et al. (2016), available as a preprint during completion of this manuscript, analyzes the structural and compositional properties of CO white dwarfs in connection with uncertainties in the H- and He-burning rates obtained by a new Monte Carlo sampling procedure.
The CO white dwarf remnant of their 3 $M_\odot$ star exhibits a C-rich outer core, which independently shows that our nova nucleosynthesis predictions apply to lower progenitor masses.
By evolving a grid of evolutionary models of 1 to 6 $M_\odot$, they find that rate uncertainties,
 especially those affecting the 
$3\alpha$ and $^{12}$C($\alpha$, $\gamma$) reactions,  
 do not have a large impact on the final white dwarf mass distribution, 
but can significantly affect the central $^{12}$C and $^{16}$O abundances. 
Since uncertainty factors are likely temperature-dependant, it is unclear how the C and O profiles in the outer core would be affected, which is the region of interest for nova nucleosynthesis.
As a conclusion,  current nuclear uncertainties do not significantly affect
the formation of C-rich layers in the outer part of CO white dwarfs,
and hence, the production of C-rich ejecta in CO nova outbursts.

It is finally worth noting that production of C-rich ejecta in nova outbursts 
may also account for the 
origin of C-rich J-type stars. These stars represent 10\% -- 15\% 
of the observed C stars in our Galaxy and in the Large Magellanic Cloud. Indeed, a binary-star formation channel for C-rich J-type 
stars has been recently proposed in the
framework of re-accretion of C-rich nova ejecta onto  main sequence companions 
(Sengupta, Izzard, \& Lau 2013).

{\it Acknowledgements.} 
We thank the referee, Robert Gehrz, for a very positive feedback on
 the manuscript.
One of us (JJ) would like to express his gratitude to the
 Center of Advanced Mathematical Sciences, American University of Beirut 
(AUB, Lebanon) for financial support and hospitality during his visit
to AUB, where part of this project started.
This work has been partially
supported by the Spanish MINECO grant AYA2014-59084-P,
 by the E.U. FEDER funds, by the AGAUR/Generalitat de Catalunya grant 
SGR0038/2014 (JJ). 
GH thanks Christopher Tout and Robert Izzard for helpful discussions, the STFC
for funding her postdoctoral research at the Institute of Astronomy,
 and the FAS at the American
University of Beirut where part of the calculations were performed.


\begin{thebibliography}{9}
\bibitem[]{Aka15} Akahori, T., Funaki, Y., \& Yabana, K. 2015, 
           Phys. Rev. C, 92, 022801 
\bibitem[]{Ama01} Amari, S., Gao, X., Nittler, L. R., et al. 2001,
           ApJ, 551, 1065
\bibitem[]{Ama02} Amari, S. 2002, NewAR, 46, 519
\bibitem[]{And11} Andersen, A. C. 2011, in {\it Why Galaxies Care about AGB 
           Stars II: Shining Examples and Common Inhabitants}, ed. 
           F. Kerschbaum, T. Lebzelter, \& R. F. Wing 
           (San Francisco, CA: Astron. Soc. Pac. Conf. Series), 215
\bibitem[]{Ang99} Angulo, C., Arnould, M., Rayet, M., et al. 1999, NPA, 656, 3 
\bibitem[]{Cas10} Casanova, J., Jos\'e, J., Garc\'\i a--Berro, E., Calder, A., 
       \& S. N. Shore 2010, A\&A, 513, L5
\bibitem[]{Cas11a} ---. 2011a, A\&A, 527, A5
\bibitem[]{Cas11b} Casanova, J., Jos\'e, J., Garc\'\i a--Berro, E., Shore, 
        S. N., \& Calder, A.~C. 2011b, Nature, 478, 490
\bibitem[]{Cla76} Clayton, D. D., \& Hoyle, F. 1976, ApJ, 203, 490
\bibitem[]{Eva90} Evans, A. 1990, in 
           {\it Physics of Classical Novae}, ed. A. Cassatella, \& R. Viotti
           (Springer-Verlag, Berlin, Germany:Springer-Verlag), 253
\bibitem[]{Eva08} Evans, A., \& Rawlings, M. C. 2008, in {\it Classical Novae} 
           (2nd Ed.), ed. M. F. Bode, \& A. Evans 
           (Cambridge, UK:Cambridge Univ. Press), 308
\bibitem[]{Fie06} Fields, C. E., Farmer, R., Petermann, I., 
          Iliadis, C., \& Timmes, F. X. 2016, ApJ, in press (arXiv: 1603.06666v1 [astro-ph.SR]). 
\bibitem[]{Fuj82} Fujimoto, M. Y. 1982, ApJ, 257, 752 
\bibitem[]{Fyn05} Fynbo, H. O. U., Diget, C. Aa., Bergmann, U. C., et 
            al. 2005, Nature, 433, 136 
\bibitem[]{Geh90} Gehrz, R.D. 1990, in 
           {\it Physics of Classical Novae}, ed. A. Cassatella, \& R. Viotti
           (Springer-Verlag, Berlin, Germany:Springer-Verlag), 138
\bibitem[]{Geh98} Gehrz, R.D., Truran, J.W., Williams, R.E., \& Starrfield, S.
           1998, PASP, 110, 3
\bibitem[]{Geh02} Gehrz, R.D. 2002, in {\it Classical Nova Explosions}, ed. 
           M. Hernanz, \& J. Jos\'e (Melville, NY:American Inst. Phys.), 198
\bibitem[]{Geh08} Gehrz, R.D. 2008, in {\it Classical Novae} 
           (2nd Ed.), ed. M. F. Bode, \& A. Evans 
           (Cambridge, UK:Cambridge Univ. Press), 167
\bibitem[]{Hae16} Haenecour, P., Floss, C., Jos\'e, J., et al. 2016, 
           ApJ, in press
\bibitem[]{Hal12} Halabi, G. M., El Eid, M. F., \& Champagne, A. 2012,
           ApJ, 761, 10
\bibitem[]{Ili15} Iliadis, C. 2015, {\it Nuclear Physics of Stars} (2nd Ed.),
           (Weinheim, Germany:Wiley-VCH Verlag)
\bibitem[]{Ili02} Iliadis, C., Champagne, A., Jos\'e, J., Starrfield, S.,
          \& Tupper, P. 2002, ApJS, 142, 105
\bibitem[]{Ish13} Ishikawa, S. 2013, Phys. Rev. C, 87, 055804 
\bibitem[]{Jos16} Jos\'e, J. 2016, {\it Stellar Explosions: Hydrodynamics 
           and Nucleosynthesis} (Boca Raton, FL:CRC/Taylor and Francis)
\bibitem[]{Jos07} Jos\'e, J., Garc\'\i a--Berro, E., Hernanz, M., \& Gil-Pons, 
           P. 2007, ApJL, 662, L103
\bibitem[]{Jos98} Jos\'e, J., \& Hernanz, M. 1998, ApJ 494, 680 
\bibitem[]{Jos04} Jos\'e, J., Hernanz, M., Amari, S., Lodders, K., \& 
           Zinner, E. 2004, ApJ, 612, 414
\bibitem[]{Jos06} Jos\'e, J., Hernanz, M., \& Iliadis, C. 2006, NPA 777, 550 
\bibitem[]{Jos08} Jos\'e, J., \& Shore, S. N.  2008, in 
           {\it Classical Novae} (2nd Ed.), ed. M. F. Bode, \& A. Evans 
           (Cambridge, UK:Cambridge Univ. Press), 121
\bibitem[]{Kar16} Karakas, A. \& Lugaro, M. 2016, ApJ, submitted.
\bibitem[]{Kat12} Katsuma, M. 2012, ApJ, 745, 192
\bibitem[]{Kel13} Kelly, K. J., Iliadis, C., Downen, L., Jos\'e, J., \& 
                  Champagne, A. 2013, ApJ, 777, 130
\bibitem[]{KP97} Kovetz, A. \& Prialnik, D. 1997, ApJ, 477, 356
\bibitem[]{Lod03} Lodders, K. 2003, ApJ, 591, 1220
\bibitem[]{Mac83} MacDonald, J. 1983, ApJ, 267, 732
\bibitem[]{Oga09} Ogata, K., Kan, M., \& Kamimura, M. 2009, Progr. Theor. 
           Phys., 122, 1055 
\bibitem[]{Pep11} Pepin, R. O., Palma, R. L., Gehrz, R. D.,
          \& Starrfield, S. 2011, ApJ, 742, 86
\bibitem[]{Sal13} Sallaska, A. L., Iliadis, C., Champange, A. E., et al. 2013,
           ApJS, 207, 18
\bibitem[]{Sed94} Sedlmayr, E. 1994, in {\it Molecules in the Stellar 
           Environment}, ed. U. G. Jorgensen (Berlin, Germany:Springer-Verlag),
           163
\bibitem[]{Sen13} Sengupta, S., Izzard, R. G., \& Lau, H. H. B. 2013,
           A\&A, 559, A66
\bibitem[]{Sha81} Shara, M. M. 1981, ApJ,  243, 926
\bibitem[]{She09} Shen, K. \& Bildsten, L. 2009, ApJ, 692, 324
\bibitem[]{Sho94} Shore, S. N., Starrfield, S., Gonzalez-Riestra, R., 
           Hauschildt, P. H., \& Sonneborn, G. 1994, Nature, 369, 539
\bibitem[]{Sho04} Shore, S. N., \& Gehrz, R. D. 2004, A\&A, 417, 695
\bibitem[]{Str39} Stratton, F. J. M., \& Manning, W. H. 1939, 
           {\it Atlas of Spectra of Nova Herculis 1934} (Cambridge, UK:Solar 
           Physics Observ.)
\bibitem[]{Xu13} Xu, Y., Takahashi, K., Goriely, S., et al. 2013, 
           NPA, 918, 61
\end{thebibliography}
\end{document}